# Colorimetric skin tone scale for improved accuracy and reduced perceptual bias of human skin tone annotations

Cynthia M. Cook[1+], John J. Howard[1+], Laura R. Rabbitt[1+], Isabelle M. Shuggi[1+*], Yevgeniy B. Sirotin[1+], Jerry L. Tipton[1+], Arun R. Vemury[2+]

[1] SAIC Identity and Data Sciences Laboratory

[2] U.S. Department of Homeland Security's Science and Technology Directorate

[+] Authors listed alphabetically.

*Corresponding author, ishuggi@idslabs.org.

Human image datasets used to develop and evaluate technology should represent the diversity of human phenotypes, including skin tone. Datasets that include skin tone information frequently rely on manual skin tone ratings based on the Fitzpatrick Skin Type (FST) or the Monk Skin Tone (MST) scales in lieu of the actual measured skin tone of the image dataset subjects. However, perceived skin tone is subject to known biases and skin tone appearance in digital images can vary substantially depending on the capture camera and environment, confounding manual ratings. Surprisingly, the relationship between skin-tone ratings and measured skin tone has not been explored. To close this research gap, we measured the relationship between skin tone ratings from existing scales (FST, MST) and skin tone values measured by a calibrated colorimeter. We also propose and assess a novel Colorimetric Skin Tone (CST) scale developed based on prior colorimetric measurements. Using experiments requiring humans to rate their own skin tone and the skin tone of subjects in images, we show that the new CST scale is more sensitive, consistent, and colorimetrically accurate. While skin tone ratings appeared to correct for some color variation across images, they introduced biases related to race and other factors. These biases must be considered before using manual skin-tone ratings in technology evaluations or for engineering decisions.

## 1 INTRODUCTION

Post-hoc annotation of skin tone is common practice in human image datasets used for facial recognition and other AI tasks because it allows practitioners to assess the skin tone diversity of their training sets, reduce model bias through training data selection, and evaluate AI models for bias with respect to skin tone. Methods to label skin tone in human image datasets fall into two general categories: automated image processing or manual human labelling using skin tone scales. Prior work has implemented color measurements from images to evaluate facial skin tone of individuals in images, ranging from simple [29, 32] to sophisticated [1, 12, 19]. However, the accuracy of automated skin tone assessments can vary based on image capture conditions (e.g., lighting, camera quality, surrounding environment) and behavioral factors (e.g., pose, facial expression, attire) which impact the color of the facial sample [17, 21].

Human assessments of facial skin tone can be perceptually biased. Human color perception is affected by individual differences between people [11, 24]. Further, the perceptual appearance of a color is dependent on other colors present in its surround [3, 26]. Distinct perceptual biases are observed in human ratings of skin tone in face images. For example, perceived skin tone lightness can be influenced by lip color; with redder lips increasing and darker lips decreasing perceived skin lightness [20].

Human ratings of facial skin tone can also be biased by race. In a psychophysical study, Levin and Banaji found that the perceived lightness of grayscale faces is influenced by perceived race [25], an effect that holds across demographically diverse observers [23]. Campbell and colleagues found that raters' self-reported gender and ethno-racial category influenced facial skin tone ratings on both text-based and palette-based scales [5]. Research using the palette-based Monk Skin Tone (MST) scale found that MST skin tone labels are systematically biased by the geographical location of the rater [33].

Skin tone rating scales can be either text-based, palette-based, or image-based. Text based scales use a set of survey questions to arrive at a category [4, 13] or use a single Likert-scale [37]. Palette-based / image-based scales present a set of color patches / images for raters to pick as the best-match to the stimulus [5, 21, 28, 33]. Except for two-dimensional palette-based scales, the primary axis of variation along all skin tone scales is skin lightness. Skin tone scales are typically used to rate facial images as representing individuals with darker or lighter skin tones; however, some research has also considered skin hue (see [35]). Because these scales are used to create post-hoc skin tone annotations, it is important to understand the degree to which scale ratings correspond to actual skin tones with a focus on skin lightness of the individual in the image.

Prior work on skin tone rating scales [5, 16, 21, 33] has not attempted to relate ratings to standardized color measurements. Indeed, most prior research has relied on the Fitzpatrick Skin Type (FST), developed as a screening for UV-sensitivity [13]. Despite prescribing precise color values for each patch [14], prior studies of palette-based scales have not related the color of the selected rating patch to the standardized color measurement of the rated face. Likewise, the impact of known biases in skin tone ratings has not been quantitatively related to variation in rated skin tone. Further, to our knowledge no prior work has tested the accuracy of a palette-based scale on self-ratings of skin tone.

Here we sought to understand the factors that influence skin tone scale preference and the relationship between scale ratings and CIELAB skin tone measurements made using a calibrated colorimeter. We implemented two survey-based studies, each with a large sample of diverse raters.

In the first study, we asked raters to rate their own skin tone using two existing scales (FST, MST), and a new Colorimetric Skin Tone (CST) scale. Self-rating of skin tone answers basic questions about such scales by removing any camera and image presentation effects which are present when rating the skin tone of other people in images. The FST and MST scales have been used in prior work assessing skin tone from facial images [4, 16, 21, 22]. We developed the new CST scale using colorimetric measurements of skin tone acquired by our group based on data collections from diverse volunteers. We examined the effect of scale type, scale presentation, and volunteer demographics on skin tone ratings and assessed the relationship between ratings and calibrated skin tone values.

In the second study, we asked raters to rate the skin tone of other individuals based on face images acquired on different imaging devices using the MST and CST scales. We examined the effect of scale type, demographics of the raters, demographics of the rated individuals, and imaging device (i.e., camera) on skin tone ratings and assessed the relationship between human ratings and calibrated skin tone values.

## 2 METHODS

### 2.1 Informed Consent

The research protocols for both studies were approved by an Institutional Review Board (IRB). All volunteers were briefed about the study and provided informed consent to participate. Volunteers were provided an opportunity to ask questions about the study in a one-on-one meeting with a researcher. Images of volunteers who consented to having their images shared in publications are presented in this work.

### 2.2 Volunteer Demographics

For both studies, data was collected at two separate biometric data collection events. A total of 1,857 volunteers participated in study 1 and a total of 1,645 volunteers participated in study 2. In study 1, eight volunteers were removed from the analysis because they chose not to have their skin tone measured with the colormeter and in study 2, 22 volunteers were removed due to rater exclusion criteria (see Section 2.6.2 – Study 2: Rating Skin Tone from Face Images). Volunteers self-reported race, ethnicity, gender, and age during the pre-study screening process. Race and ethnicity were combined into ethno-racial categories following conventions in the 2020 Census Diversity Index [31]: Hispanic or Latino; White, not Hispanic or Latino; Black or African American, not Hispanic or Latino; Asian, not Hispanic or Latino; and Other. Volunteers who selected Other as their race and/or did not specify their gender as Female or Male were removed from the analysis due to a small sample size (study 1: 102 volunteers; study 2: 100 volunteers).

For brevity, we will use "race" to refer to volunteers' self-reported ethno-racial category. Overall, in the first study 5.92% volunteers were removed and in the second study 7.42% of volunteers were removed. Figure 1 shows the demographics of volunteers by race, gender, age, and calibrated skin tone metrics that were included in study 1.

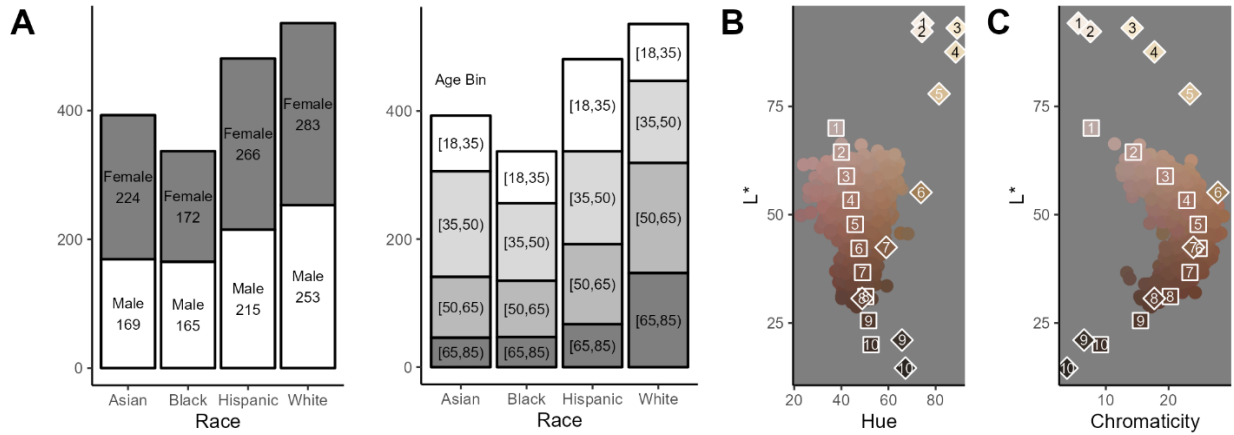

Figure 1. Study 1 Demographics. A) Counts of volunteers by race, gender, and age. Left: dark gray bars show count of Females and white bars show count of Males. Right: age bins are reported ranging from age bin of [18, 35) in white to age bin of [65, 85) in dark gray. B) Hue and lightness of each volunteer overlapped with the hue and lightness from each color-based scale. C) Chromaticity and lightness of each volunteer overlapped with the chromaticity and lightness from each color-based scale. In both panels B and C, the squares represent the CST scale, and the diamonds represent the MST scale.

Figure 2 shows the demographics of volunteers that participated in study 2 by race, gender, and age. The skin tone metrics of volunteers were not relevant for study 2 and are not presented.

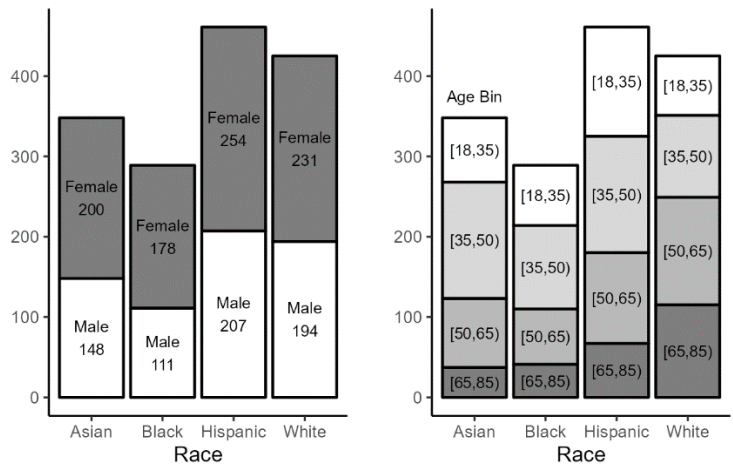

Figure 2. Study 2 Demographics. Counts of volunteers by race, gender, and age. Left: dark gray bars show count of Females and white bars show count of Males. Right: age bins are reported ranging from age bin of [18, 35) in white to age bin of [65, 85) in dark gray.

### 2.3 Colorimetric Skin Tone Measurements

Skin tone was measured using a calibrated hand-held sensor (DSM III Colormeter, Cortex Technology). The sensor measures skin color using an RGB sensor to image a 7 mm$^2$ patch of skin under standard illumination provided by two white light emitting diodes. The device has been shown to accurately measure the color of human skin [6, 9].

For each of the volunteers, two bilateral measurements were collected. First, bilateral hand skin tone measurements were collected from skin covering the first interosseus muscle. Second, bilateral facial skin tone measurements were collected from skin covering the zygomatic arch. The four sRGB measurements were collected in close succession and converted to the CIELAB color space using the D65 illuminant. The subjects' skin was not cleaned prior to collection. The skin contacting surfaces of the colormeter were wiped with rubbing alcohol between subjects and the device itself was calibrated twice a day using a standardized procedure involving a white calibration plate provided by the colormeter manufacturer. We verified that our ground-truth skin tone readings matched skin tone readings reported in prior work [27, 36, 38]. The CIE $L^*$ $a^*$ $b^*$ values from the colormeter for the bilateral hand and face measurements were averaged to obtain a single face and hand skin tone for each volunteer. In this study, skin tone is described in terms of average lightness ($L^*$), hue, and chromaticity as they have superior interpretability and align better with human perceptual experience of color and neurophysiological organization of the visual cortex [25]. Hue and chromaticity were calculated from average $a^*$ and average $b^*$ as follows.

$$Hue = \frac{180}{\pi} atan\left(\frac{b^*}{a^*}\right)$$

$$Chromaticity = \sqrt{a^{*2} + b^{*2}}$$

## 2.4 Expected Minimum Human Error

Expected minimum for human error in skin tone rating was estimated based on the distance in color space between each bilateral measurement. We assume that human error cannot be less than the error found between bilateral measurements using a calibrated hand-held sensor as follows:

$$\Delta E_{i,rl} = \sqrt{\left(L_{i,r}^* - L_{i,l}^*\right)^2 + \left(a_{i,r}^* - a_{i,l}^*\right)^2 + \left(b_{i,r}^* - b_{i,l}^*\right)^2}$$

$$\Delta E_{min} = \frac{1}{N} \sum_i \Delta E_{i,rl}$$

Where $\left(L_{i,r}^*, a_{i,r}^*, b_{i,r}^*\right)$ is the skin tone measured on the right location and $\left(L_{i,l}^*, a_{i,l}^*, b_{i,l}^*\right)$ is the skin tone measured on the left location for subject $i$, $N$ is the number of volunteers, $\Delta E_{i,rl}$ is the color difference and $\Delta E_{min}$ is the minimum expected human error. Figure 3 shows the distribution of $\Delta E_{i,rl}$ and $\Delta E_{min}$ of bilateral hand (A) and temple (B) measurements. Because $\Delta E_{min}$ values for hands and temples were similar, we use the larger $\Delta E_{min}$ value of 3.5 as the expected minimum human error throughout.

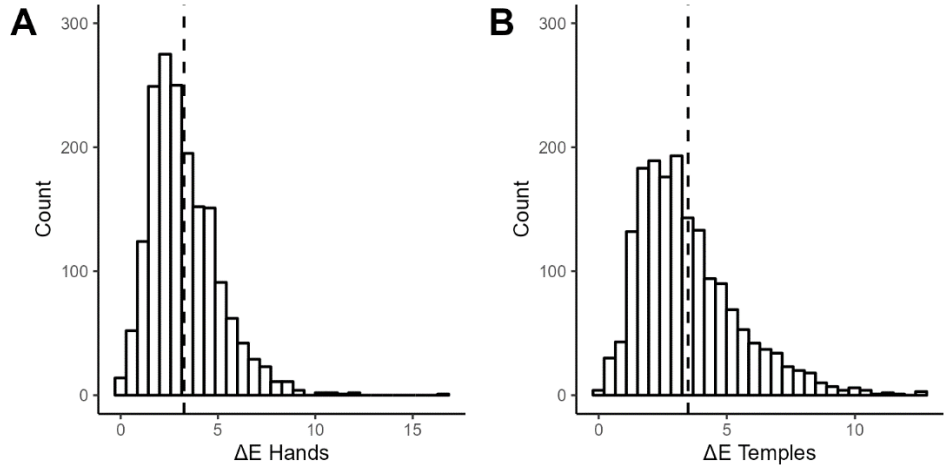

Figure 3. Distribution of $\Delta E_{i,rl}$ for bilateral hand (A) and temple (B) measurements. The mean $\Delta E_{min}$ is indicated by the dashed line in each panel (hands = 3.3, temples = 3.5).

## 2.5 Skin Tone Rating Scales

Three rating scales were examined. The FST scale is a six-point text-based scale originally developed to predict UV-sensitivity of different skin types [12] and has been used in prior work as a skin "color" scale [4]. The FST scale questions were adapted from [10]. It was used in study 1 as a text-based comparison for two palette-based scales, the MST scale and the CST scale (Figure 4).

The MST is a ten-point scale (Figure 4B), which has been used in prior work and was developed to capture ethno-racial diversity in skin tone across North and South America [16, 30, 33]. The methodology used to develop the 10 colors used in the MST palette is not publicly known. In addition to capturing diversity in skin tones, the scale is intended to dissociate skin tone and race through two different formats [14]. The first format is referred to as orbs, which are colored circles meant to capture differences in each MST group by shading each orb to show variations in skin tone. The second format is ten colored rectangles of the most prominent color in the orbs, referred to as swatches. The orbs are meant to be used for annotations of an image dataset or volunteers in a study, while the swatches are meant for researchers who need to use exact color values. Our work required use of the swatches because the exact colors were required for comparison to the calibrated skin tone measurements.

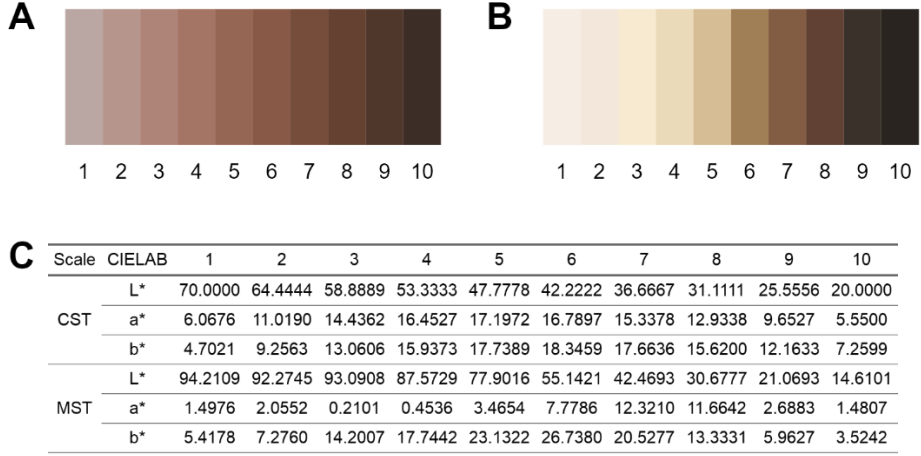

Figure 4. A) CST and B) MST scales presented on a white background. C) CIELAB values for each color swatch on each scale.

The CST scale was developed based on 2,517 calibrated facial skin tone measurements (see Calibrated Skin Tone Measurements) collected in prior studies conducted by our laboratory. The CST scale was created by estimating hue and chromaticity at evenly spaced increments of 10 $L^*$ values from 20 to 70 resulting in the colors shown in Figure 4A. Hue and chromaticity were estimated as a function of $L^*$ using linear regression such that:

$$\widehat{Hue}(L^*) = \hat{\beta}_o + \hat{\beta}_1 L^* + \hat{\beta}_2 L^{*2}$$

$$\widehat{Chromaticity}(L^*) = \hat{\beta}_o + \hat{\beta}_1 L^* + \hat{\beta}_2 L^{*2}$$

Prior work has defined six skin tone types in the CIELAB color space using the Individual Typology Angle (ITA) [8]. However, ITA calculation discards the $a^*$ color component and assumes a fixed reference $L^*$ value of 50 without theoretical support. The ITA types for CST values are as follows: 1-2 are "very light", 3 is "intermediate", 4 is "tan", 5-6 are "brown", and 7-10 are "dark".

## 2.6 Rating Tasks and Analyses

All skin tone rating tasks were presented as surveys on tablet computers (Apple iPad A1893) at two distinct test locations. No attempt was made to calibrate tablet screens. Most applied research employing human skin tone annotation does not include screen calibration as it relies on raters performing the task in a variety of remote locations [21, 30, 33]. We refer to volunteers who provided skin tone ratings as raters.

### 2.6.1 Study 1: Rating Own Skin Tone

Each palette-based scale (Figure 4) was presented to the raters with the following instruction: "Using the scale below, select the number corresponding to the color that you think best matches your skin tone."

**Color accuracy:** Accuracy of self-rating was based on the distance in CIELAB color space ($\Delta E$) between the color $(L_{i,s}^{*,swatch}, a_{i,s}^{*,swatch}, b_{i,s}^{*,swatch})$ of swatch ($i$) on scale ($s$) and the average measured skin tone $(L_{i,s}^{*,avg}, a_{i,s}^{*,avg}, b_{i,s}^{*,avg})$ of the raters hands choosing this swatch:

$$\Delta E_{i,s} = \sqrt{(L_{i,s}^{*,avg} - L_{i,s}^{*,swatch})^2 + (a_{i,s}^{*,avg} - a_{i,s}^{*,swatch})^2 + (b_{i,s}^{*,avg} - b_{i,s}^{*,swatch})^2}$$

**Response modeling:** For each scale (MST, CST, and FST) we used multiple linear regression to model responses based on skin lightness, hue, chromaticity, race, gender, background color, and location of data collection (the equation below shows the full model used for each scale). Additionally, location was added to the model as a control variable for differences across the two samples. All continuous variables were mean centered.

$$response_i = \beta_0 + \beta_1 lightness_i + \beta_2 hue_i + \beta_3 chromaticity_i + \beta_4 race_i + \beta_5 gender_i + \beta_6 background_i + \beta_7 location_i + \varepsilon_i$$

The optimal model for each scale was selected using stepwise Bayesian Information Criterion (BIC), $BIC = k \ln(n) - 2 \ln(\hat{L})$, where $k$ represents the number of estimated parameters in the model and $\hat{L}$ represents the maximum value of the model's likelihood function. We applied the stepwise procedure in both directions using the `step()` function in the R package `stats`. To determine the relationship between lightness and other variables of interest, we calculated the $L^*$ ratio ($L^* \, ratio_x = \beta_x / \beta_1$) using the coefficients of each variable ($\beta_x$) and the coefficient of lightness ($\beta_1$) for each model.

### 2.6.2 Study 2: Rating Skin Tone from Face Images

Each face image (Figure 5) was presented to the raters on one of the palette-based scales (Figure 4) with the following instruction "Which item in the scale corresponds best to the complexion of the individual pictured below?"

**Selection of Subjects and Images**: Eight subjects were selected to be presented to the raters as part of study 2 from a previous data collection completed by our group. Those subjects were balanced across gender (self-identified as female or male), race (self-identified as Black or White; ethnicity was not considered as part of this selection criteria), and $L^*$ as measured by the colormeter (minimum and maximum). For each subject, images captured by three devices were selected to include a range of image exposures (see Figure 5).

**Rater exclusion criteria:** Raters completed two attentional questions, based on the 10-point scale they had used to rate each subject's skin tone. Each attentional question asked them to match the presented color swatch to the corresponding color on the scale they were using to rate images (i.e., the rater needed to match color swatch 4 on either scale within ±1 of the number 4 response on the scale). As previously mentioned, 22 raters did not meet this criterion and were removed from further analysis. Additional data cleaning was completed by checking for outliers on each question requiring skin tone ratings of subjects (94 of the 12,184 collected responses were removed).

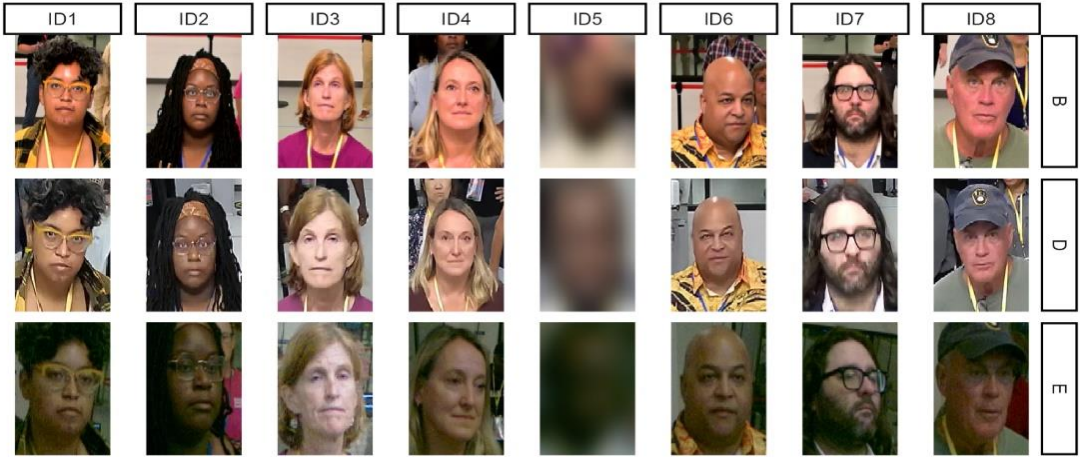

Figure 5. Images presented in study 2. Raters randomly rated one image from each subject. ID5 did not consent to images being used in publications and therefore is blurred.

**Rater accuracy:** Rater accuracy in study 2 was assessed following the same method outlined for study 1, however, for study 2, the averaged values used were of the subjects' faces in the images for each selected response on each scale $\left(L_{i,s}^{*,avg}, a_{i,s}^{*,avg}, b_{i,s}^{*,avg}\right)$. Assessment of accuracy was collapsed across all three devices. To assess the consistency of responses for each device on each scale, the intraclass correlation was calculated using the ICC() function in the R package psych. Specifically, a two-way random effects model was used.

**Response modeling:** We modeled the ratings in study 2 using linear mixed effect modeling, which allowed us to determine the degree to which our identified predictors of interest impacted ratings, while properly accounting for the variance due to both the images being rated and the raters themselves. Specifically, we modelled the ratings, $response_{i,j}$, as a function of the image properties: lightness, hue, chromaticity, gender, race, and device and rater properties: race and gender. Since, we hypothesized that the identity of the subject could also have an impact on rating (i.e., each subject would have a slightly different optimal model), we included a random intercept for each subject, $b_{0,i}$. Notably, the variance in the random intercept terms was essentially 0, meaning the fixed effects in our model account for additional variation due to subject.

The equation below shows the linear mixed effect model for each scale; unlike study 1, location was not necessary as a control variable since the task required rating of others skin tones. All continuous variables were mean centered.

$$response_{i,j} = \beta_0 + \beta_1 lightness_i + \beta_2 hue_i + \beta_3 chromaticity_i + \beta_4 subject.race_i \\ + \beta_5 subject.gender_i + \beta_6 device_i + \beta_7 rater.race_j + \beta_8 rater.gender_j + b_{0i} + \epsilon_{i,j}$$

Where *i* represents the unique image ($1 \leq i \leq 24$) and *j* represents the unique rater who rated the image ($1 \leq j \leq 1523$). The *L\** ratio was calculated for each model as previously described (see Study 1: Rating own skin tone, Response modeling).

# 3 RESULTS

## 3.1 Study 1: Rating Own Skin Tone

To determine raters' accuracy when self-rating skin tone, their preference between two palette-based scales, and which demographic and presentational factors impact self-rating of skin tone, we asked raters to rate their own skin tone using the FST, MST and CST scales. For the FST scale, raters were instructed to determine which of the following descriptions best matches their skin type. When presented with the MST or CST scales, raters were asked to select the number corresponding to the color they thought best matched their skin tone. The MST and CST scales were randomly presented on either a white (N = 877) or gray (N = 870) background. The order of MST and CST scale presentation was randomized for each rater, the FST scale was presented at the end of the survey.

### 3.1.1 Scale Color Accuracy

Color accuracy of self-rating was assessed using CIELAB space color distance (see Section 2.6. – Rating Tasks and Analyses). Figure 6 shows the color error (ΔE) for individual skin tone swatches (A) and the average color error across all swatches disaggregated by the raters' race (B). The overall color error of scale responses was generally higher than the expected minimum possible human error (see Section 2.4 - Expected Minimum Human Error; $\Delta E_{min} = 3.5$). Raters were more accurate when using the CST scale, showing the highest accuracy for responses in the middle of the scale. All but two swatches of the CST scale presented on the white background achieved a color error value below 15 ($\Delta E_{i,CST,white} < 15, i \in \{2,3,\ldots,9\}$). In comparison, on the MST scale for both background colors only three swatches had equally low error ($\Delta E_{i,MST,white} < 15, i \in \{6,7,9\}$). Notably, on the MST scale all responses selected for swatches 1-5 (lighter skin tones) resulted in higher error ($\Delta E_{i,CST,b} < \Delta E_{i,MST,b}, i \in \{1,2,\ldots,5\}, b \in \{white, gray\}$). Disaggregated by race, raters who self-identified as Black were more accurate (lower color error) compared to other raters on the MST scale. The CST scale maintained similar levels of accuracy across all race groups (Figure 6B).

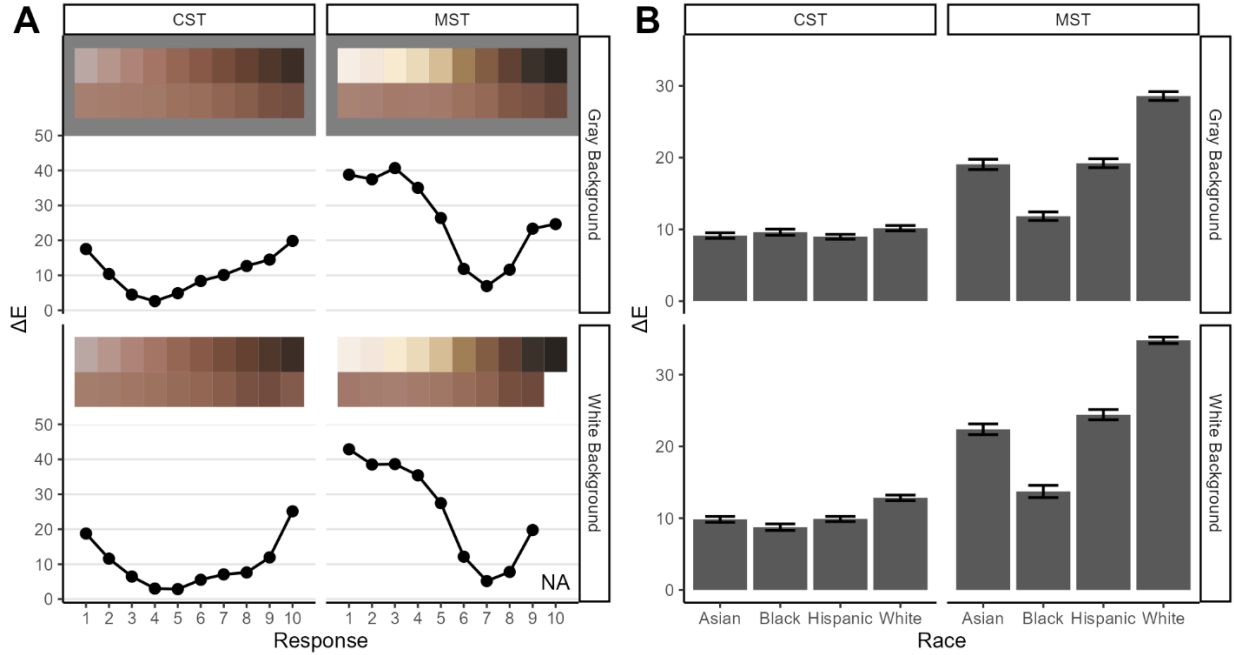

Figure 6. A) Accuracy of self-rating skin tone for CST and MST scales, separated by background color; within each panel the top color-palette shows the relevant scale, and the bottom color-palette is the average color of raters who selected that response. B) Accuracy of self-rating skin tone for CST and MST scales disaggregated by self-identified race.

### 3.1.2 Scale Preference

After the MST and CST rating tasks, raters were asked whether the MST or CST scale had a better match to their skin tone. We found that the raters' race as well as the background on which scale was presented affected scale preference (Figure 7A). On a gray background, the CST scale was preferred by raters of all race groups. Preference for the CST scale was greatest for raters who identified as Black and lowest for those who self-identified as White. On a white background, Black raters continued to favor the CST scale, however, Asian, Hispanic, and White raters now favored the MST scale. Interestingly, we found that scale preference was also systematically related to rater skin lightness (Figure 7C) such that raters with progressively lighter skin tone favored the MST scale relative to the CST scale. Preference was modelled using an optimal logistic regression which included background color and hand skin lightness (optimal model selection was completed as described in Section 3.1 - Study1: Rating own skin tone).

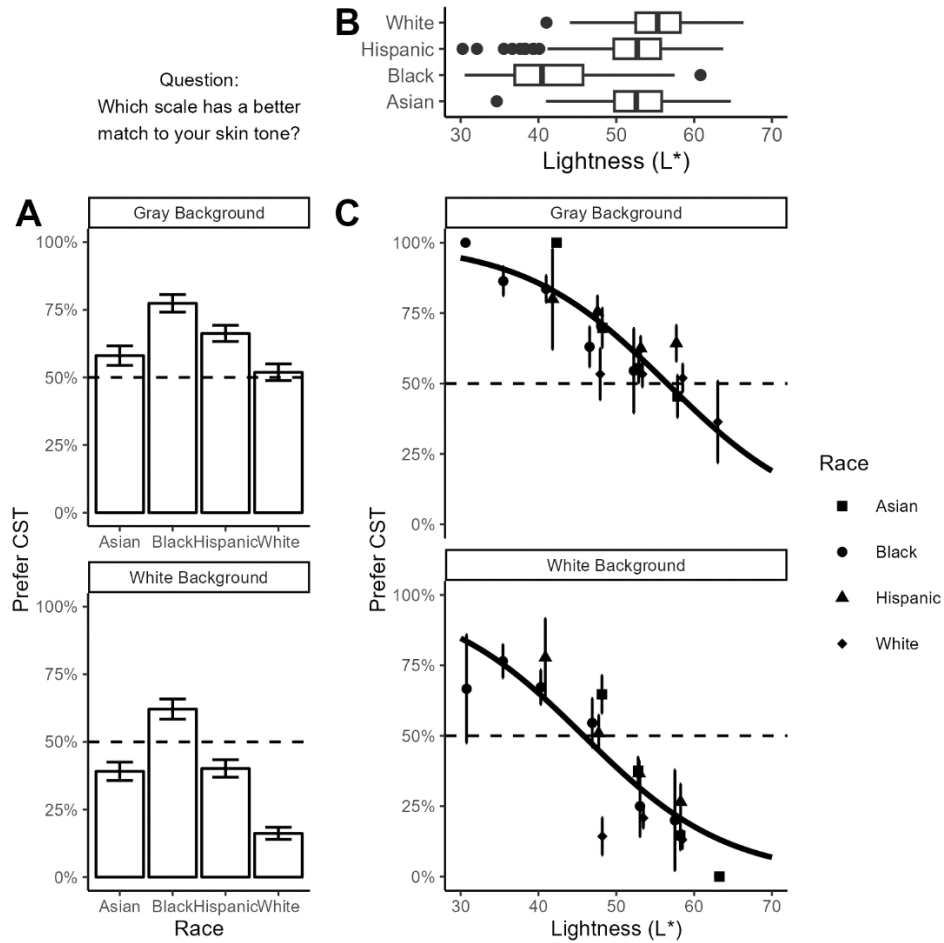

Figure 7. A) Rater preference by background color and race, percentage represents raters who selected the CST scale as the best match for their skin tone. B) Skin lightness values of raters by race. C) Percentage of raters who preferred the CST scale on each background color plotted by binned skin lightness and race.

### 3.1.3 Factors Affecting Rating Scale Responses

We examined how rater responses on the two palette-based rating scales and the FST scale were related to their calibrated skin tone measurements (see Section 2.3 - Colorimetric Skin Tone Measurements) as well as other factors that may bias scale ratings. Figure 8 visualizes the relationship between raters' chosen scale responses, their skin lightness, and race.

As expected, all scale ratings correlated with skin lightness. However, scales differed based on the utilization of available scale range. Average responses utilized 66% of the available scale range for the text-based Fitzpatrick skin type (FST). Palette-based scales utilized more of the available scale range. Average CST ratings utilized 80%-90% of the available scale range whereas average MST ratings utilized 70%-80% of the available scale range. Interestingly, on the palette-based scales, raters who self-identified as White selected systematically lighter swatches as compared to raters of other races that had similar measured skin tone values (i.e., within the same $L^*$ bin).

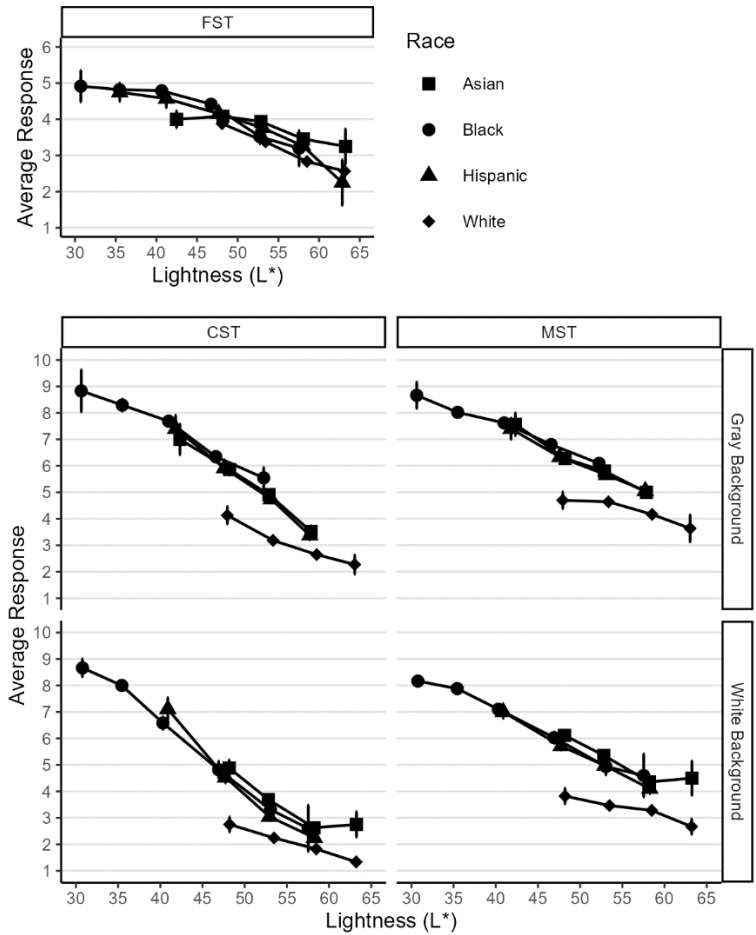

Figure 8. Relationship between raters' average skin tone scale ratings, their skin lightness, and race. Measured rater skin lightness was split into equally spaced L* bins. The average L* value of raters in each bin is plotted against the average scale rating (on the CST and MST scales, 1 represents the lightest skin tone and 10 represents the darkest skin tone, see Figure 4 for scales). Top: average FST responses for raters with skin lightness falling in each L* bin. Bottom: CST and MST responses separated by background color. Error bars are SEM.

We used linear regression and optimal model selection to examine the factors that influenced scale responses (see Rating Tasks and Analyses). Modeling started with an initial full model considering factors related to rater skin tone (lightness, hue, chromaticity), rater demographics (gender, race), as well as scale background and the control factor of test location. Model selection produced an "optimal" model which included only those factors that improved model fit relative to the number of model parameters.

The optimal model for FST ratings, included factors related to rater skin tone (lightness, hue, chromaticity) and race (adjusted $R^2 = 0.30$, $F(6, 1740) = 128.6$, $p < 0.001$). The fits of the optimal models for the palette-based scale responses were notably better. For CST ratings (adjusted $R^2 = 0.61$, $F(7, 1739) = 390$, $p < 0.001$) and MST ratings (adjusted $R^2 = 0.57$, $F(7, 1739) = 332.8$, $p < 0.001$) the same optimal model was selected. The optimal model included factors related to rater skin tone (lightness, hue, chromaticity), race, scale background, as well as test location. Adjusted $R^2$ values for optimal palette-based scale rating models were roughly two-fold higher than for the FST with the optimal CST scale rating model showing a somewhat better fit relative to that of the MST.

Of the three factors related to rater skin tone, skin lightness ($L^*$) was uniformly most strongly related to the scale responses with higher $L^*$ associated with lower scale ratings. The full range of rater skin lightness spanned $L^*$ values between 30 and 70 (Figure 1). The slope of $L^*$ with scale ratings showed that each FST rating step was equivalent to 14.7 *L\** units whereas each step along the CST scale was equivalent to just 4.9 L\* units and each step along the MST scale was equivalent to 7.4 *L\** units. The size of the relationship between scale rating and *L\** corresponds to the sensitivity of the ratings to *L\** with smaller values indicating greater sensitivity. Correcting by the number of scale responses, the CST scale had the highest sensitivity to *L\**, followed by the MST, and then FST.

Hue was selected in all optimal models suggesting that skin hue contributes to scale response selection. The full range of rater skin hue spanned values between 20 deg and 65 deg (Figure 1). The slopes of hue with scale ratings showed that 52 deg, 35 deg and 32 deg of hue were equivalent to one step increase in FST, CST, and MST ratings, respectively (Table 1). Thus, none of the scales were very sensitive to hue, with the largest possible hue difference in the population (about 45 degrees) corresponding to little more than a one-step difference in scale response for the MST and CST scales. Relative to lightness, hue was least related to scale rating on the CST scale and most on FST (Table 1, L\* Ratio). Lower (pinker) hue values affected scale ratings similarly to increases in *L\**, like a previously observed perceptual relationship between skin hue and lightness [15, 20].

For the palette-based scales, background color and location were both significant predictors of response according to the optimal models. Raters who were presented with the scale on a white background reduced their ratings by 1.11 steps along the CST scale and by 0.7 steps along the MST scale on average relative to ratings on a gray background. This effect of scale background was equivalent to 5.5 *L\** units on the CST scale and 5.2 *L\** units on the MST scale and is consistent with expectations based on simultaneous brightness contrast [34]. Likewise, the effect of test location corresponded to 2.7 and 2.8 *L\** units on the CST and MST scales, respectively (Table 1, L\* Ratio). These large effects show that human skin tone ratings are strongly affected by the context within which the scale is presented.

Rater's race was selected in all optimal models. Raters who self-identified as White rated their own skin as significantly lighter compared with raters of other races, given they had the same measured skin lightness (Table 1). On average, raters who self-identified as White, reduced their scale ratings by amounts equivalent to 4.7, 6.0, and 8.6 *L\** units compared to those that self-identified as Black on the FST, CST, and MST scales, respectively. Relative to skin lightness, the MST scale appeared to be the most affected by the rater's race.

Table 1. Study 1 Multiple Linear Regression Results for Each Scale

| Scale | Covariate | Estimate | Standard Error | t-statistic | p-value | L* Ratio | Adjusted $R^2$ |
|---|---|---|---|---|---|---|---|
| FST | Calibrated L* | -0.0680 | 0.0053 | -12.8515 | 3.53e-36 | 1.0000 | 0.3048 |
| | Hue | 0.0193 | 0.0041 | 4.6536 | 3.51e-06 | -0.2832 | |
| | Chromaticity | 0.0446 | 0.0100 | 4.4541 | 8.96e-06 | -0.6558 | |
| | Asian | 0.3033 | 0.0680 | 4.4600 | 8.72e-06 | -4.4586 | |
| | Black | 0.3196 | 0.0954 | 3.3506 | 8.24e-04 | -4.6996 | |
| | Hispanic | 0.2352 | 0.0634 | 3.7103 | 2.14e-04 | -3.4583 | |
| CST | Calibrated L* | -0.2027 | 0.0075 | -26.8745 | 2.40e-133 | 1.0000 | 0.6093 |
| | Hue | 0.0287 | 0.0063 | 4.5422 | 5.95e-06 | -0.1414 | |
| | Asian | 1.1802 | 0.1069 | 11.0388 | 1.97e-27 | -5.8235 | |
| | Black | 1.2237 | 0.1546 | 7.9153 | 4.35e-15 | -6.0380 | |
| | Hispanic | 0.9801 | 0.1007 | 9.7288 | 8.09e-22 | -4.8360 | |
| | Background: White | -1.1185 | 0.0709 | -15.7743 | 1.70e-52 | 5.5188 | |
| | Location: CA | -0.5409 | 0.0892 | -6.0661 | 1.60e-09 | 2.6689 | |
| MST | Calibrated L* | -0.1360 | 0.0060 | -22.7929 | 7.79e-101 | 1.0000 | 0.5709 |
| | Hue | 0.0308 | 0.0050 | 6.1597 | 9.03e-10 | -0.2261 | |
| | Asian | 1.2161 | 0.0846 | 14.3753 | 2.32e-44 | -8.9417 | |
| | Black | 1.1704 | 0.1223 | 9.5683 | 3.57e-21 | -8.6058 | |
| | Hispanic | 1.0519 | 0.0797 | 13.1965 | 5.75e-38 | -7.7345 | |
| | Background: White | -0.7026 | 0.0561 | -12.5238 | 1.62e-34 | 5.1662 | |
| | Location: CA | -0.3752 | 0.0706 | -5.3178 | 1.19e-07 | 2.7587 | |

## 3.2 Study 2: Rating Skin Tone from Face Images

Next, we assessed the accuracy and consistency of palette-based skin tone scales when used to rate face images of other people as well as to the factors impacting these ratings. We selected images of eight subjects for rating. To investigate the effect of capture device on skin tone ratings [17], we selected three previously collected images for each subject. Each image was captured using a different imaging device (i.e., camera) on the same day (see Section 2.6.2 – Rating Skin Tone from Face Images, Selection of Subjects and Images). We asked raters to rate the skin tone of the eight subjects, based on these images using either the MST or the CST scale. Raters were shown a single image of each subject from a randomly selected device along with a palette-based scale presented on a gray background. Subjects were presented in a random order to each rater. Raters were asked to select the scale value that best matched the skin tone of the subject (Figure 4). Half of the raters (N = 760) used the MST scale to rate the skin tone of the subject and the remaining raters used the CST scale (N = 763). Some raters were removed from analysis based on exclusion criteria (see Section 2.6.2 – Rating Skin Tone from Face Images, Rater Exclusion Criteria).

### 3.2.1 Scale Color Accuracy

The accuracy of scale color ratings was calculated relative to the previously acquired calibrated skin tone color values of each of the eight subjects. Interestingly, the color accuracy from rating face images was comparable to self-rating accuracy (compare Figure 9A to Figure 6A). As before, CST scale ratings were more accurate compared to the MST scale (Figure 9A). Color error of the CST scale ratings (ΔE) was consistently lower than 15 across all responses ($\Delta E_{i,CST} < 15, i \in \{1, 2, \ldots, 10\}$). Less than half of the MST scale rating responses maintained similarly low color error ($\Delta E_{i,MST} < 15, i \in \{6, 7, 8, 9\}$). Notably, the lowest color error on the MST aligned with darker color swatches.

Figure 9B shows rating color error separately for each of the eight rated subjects. For the CST scale, color error is relatively uniform across all subjects. Color error on the MST scale was greater for subjects ID3, ID4, ID7, and ID8, all of whom self-identified as White. However, color error was not higher for subjects ID6 and ID1, who self-identified as Black but had skin lightness comparable to ID8 and ID4, respectively (see Figure 11A, Calibrated $L^*$ values).

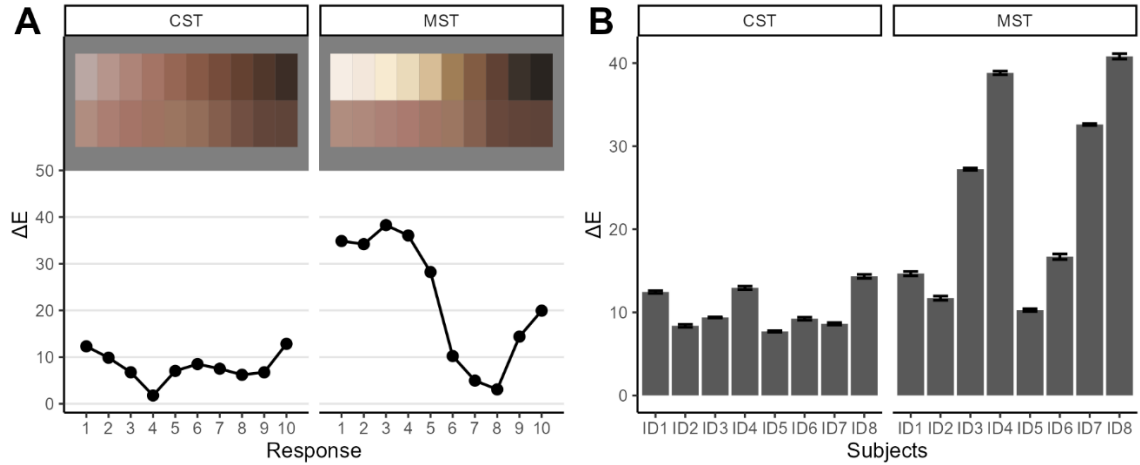

Figure 9. For each scale: A) Accuracy of image ratings averaged by selected response; within each panel the top color-palette shows the relevant scale, and the bottom color-palette is the average color of the subjects' faces rated with the corresponding response. B) Accuracy disaggregated by identity for each scale.

### 3.2.2 Scale Rating Consistency

We assessed the consistency of scale ratings by measuring the intra-class correlation coefficient (ICC) of ratings for each subject (class). We measured ICC separately for ratings obtained on different scales and for images from different imaging devices (Table 2). An ICC of 0 indicates no consistency in scale ratings across subjects for each device and an ICC of 1 indicates identical scale ratings across subjects for each device. Overall, ICC values were high (ICC > 0.8), however, the agreement in subject ratings for device B images were notably lower, especially with the MST scale. Indeed, scale ratings for device B were notably less consistent with the MST scale (ICC = 0.81) relative to the CST scale (ICC = 0.90). We found that MST scale ratings were overall less consistent than CST scale ratings with the highest MST consistency observed for device E (ICC = 0.90) comparable to the lowest CST consistency observed for device B.

Table 2. ICC for CST and MST Scales by Device

| Scale | Device | ICC |
|---|---|---|
| CST | B | 0.8986 |
| | D | 0.9214 |
| | E | 0.9185 |
| MST | B | 0.8104 |
| | D | 0.8746 |
| | E | 0.8941 |

### 3.2.3 Factors Affecting Rating Scale Responses

The skin tones of the rated subjects spanned the full range of human skin lightness values observed by our group (see Figure 1). Ratings systematically increased with $L^*$, and, similar to previous observations for self-rating, average CST scale responses spanned more of the scale range (~100%) as compared to MST (~ 80%; Figure 10).

On both scales, we found that ratings clearly varied based on the self-identified race of the subject in the image. At similar $L^*$ values, ratings of subjects who identified as Black were systematically higher (darker) than ratings of subjects who identified as White across all imaging devices (Figure 10). This resulted in a divide such that all images of subjects who identified as White were rated below 4 on the CST scale and all images of subjects who identified as Black were rated above 4. The same divide by race was observed on the MST scale, but around the value 5.

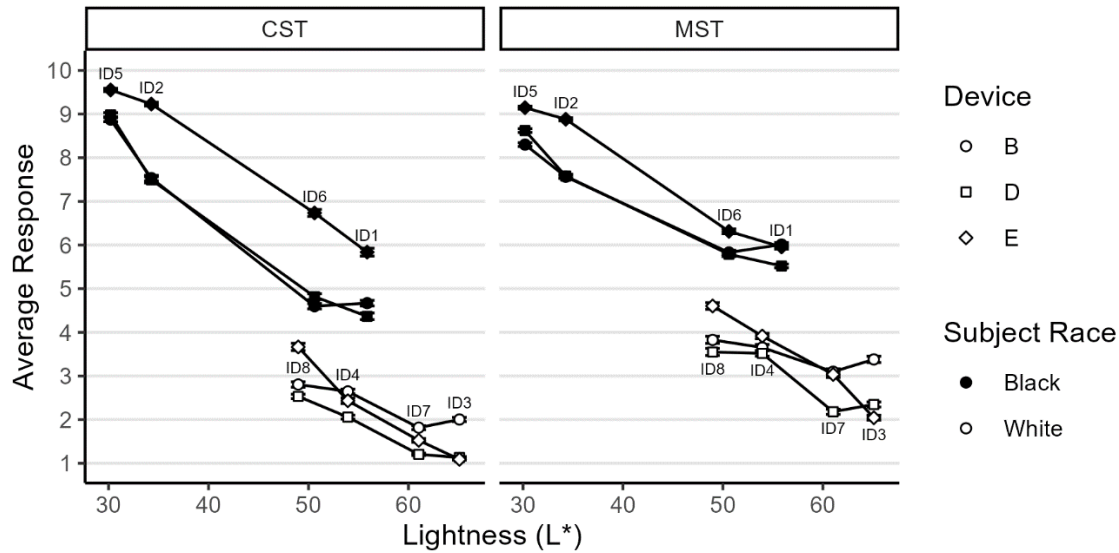

Figure 10. Average scale ratings of subjects in images versus calibrated skin lightness measurements for the subjects made using a colormeter. Note y-axis values align with scales, where 1 is the lightest and 10 is the darkest skin tone.

Interestingly, this relationship between the race of the individual in the image and rating was observed independent of a rater's race. That is, Asian, Black, White, and Hispanic raters all rated White subjects as lighter than Black subjects based on their images independent of capture device even when the subjects had the same measured skin lightness (Figure 11B; compare ratings for ID8 with those for ID6 and ratings of ID4 with ID1). This marked separation in ratings cannot be explained by the lightness of the face as reproduced in the image since the $L^*$ values measured from images do not show the same separation by race (Figure 11A).

In addition to the observed effect of the subject's race, responses on each scale were also impacted by the rater's race. Notably, raters who self-identified as Black, rated all images of ID1 and ID6 (subjects that self-identified as Black who had the highest lightness values in that race group) as having lighter skin tones compared with raters of other races (Figure 11B).

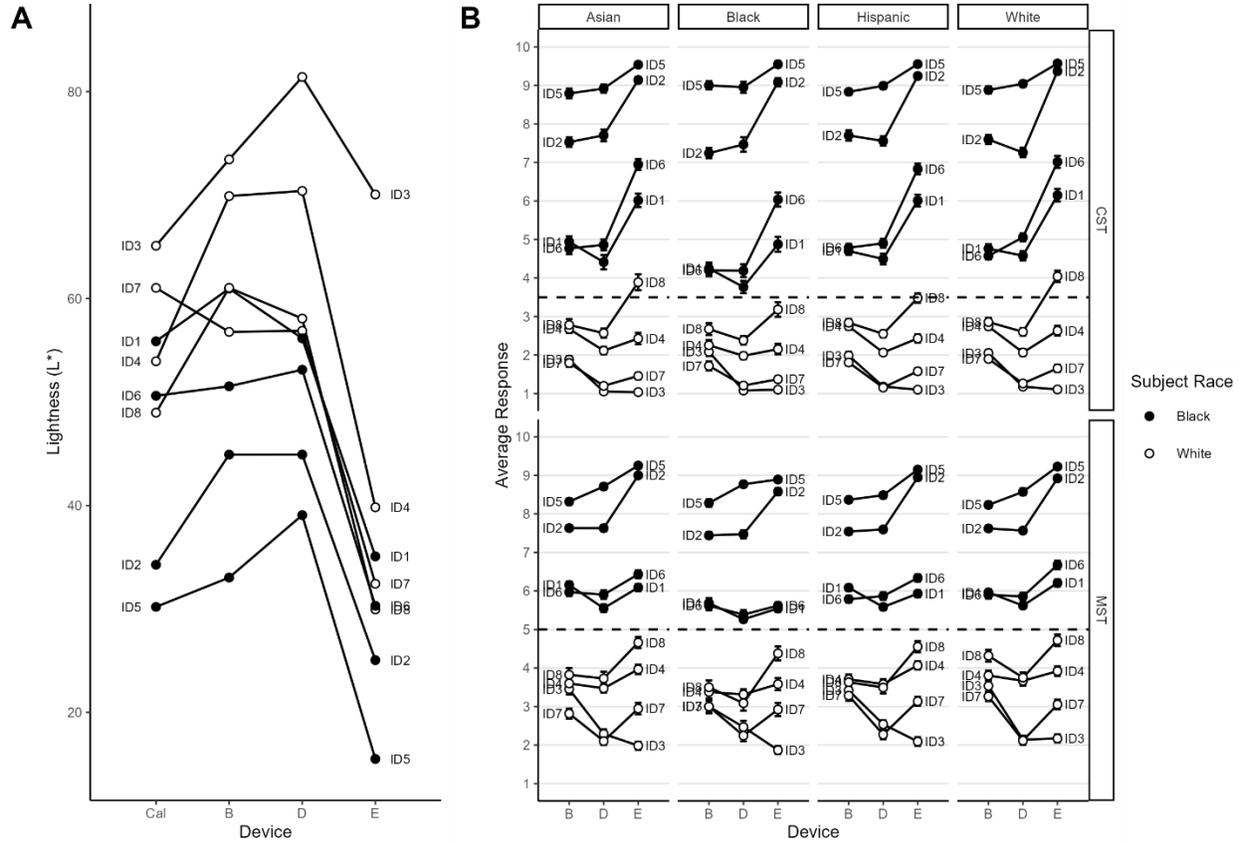

Figure 11. A) Lightness values for each subject from the calibrated color meter (Cal) and based on images acquired on different biometric devices (B, D, E). B) Average responses on each scale based on images acquired on different biometric devices. Panels show ratings by rater race. For both A and B, open and closed points denote the race of the subjects in images.

We used linear mixed-effect models to examine the factors that influenced scale responses. The mixed effect models for the CST and MST scales both resulted in a good fit to the data (conditional $R^2$ CST = 0.89; conditional $R^2$ MST = 0.84). The models included factors related to the image being rated (subject race, gender, skin lightness, hue, and chromaticity as well as image device) and the rater (race and gender). Apart from hue for MST, all factors included in the model had narrow confidence intervals and did not include zero (Table 3).

Actual skin lightness of the subject ($L^*$) was inversely related to the rater scale responses with higher $L^*$ associated with lower scale ratings. The slope of $L^*$ with scale ratings showed that each CST rating step was equivalent to 6.1 $L^*$ units and each step along the MST scale was equivalent to 8.5 $L^*$ units. This relationship between lightness and scale ratings is comparable to self-rating using the same scales albeit with somewhat lower sensitivity to $L^*$. Again, the CST scale showed greater sensitivity to skin lightness than the MST scale.

We found that the imaging device affected scale ratings. The lightness of the face images collected on different devices varied. Face image lightness measured on device E was on average 21.7 $L^*$ units lower than device B (Figure 11A). On the other hand, face image lightness measured on device D was on average 1.1 $L^*$ units higher than on device B. Compared to device B, ratings from images taken on device D were lower and those taken on device E were higher. The magnitude of this effect was comparable to an increase of 1.8 and 2.7 $L^*$ units for device D and a decrease of 3.9 and 2.4 $L^*$ units for device E on the CST and MST scales respectively, but notably different than expectations based on face image lightness (Table 3).

We observed a strong effect of the rated subjects' race on scale ratings such that subjects who identified as White were assigned much lower scale ratings relative to subjects who self-identified as Black (2.2 and 2.1 steps lower on the CST and MST scales, respectively). The magnitude of the subject race effect corresponded to 13.3 $L^*$ units on the CST scale and 17.6 $L^*$ units on the MST scale (Table 3, L* Ratio). This effect bore some similarity to the effect of race on self-ratings but was 2 to 3-fold larger (Table 1).

The race of the rater also had a sizeable effect on the ratings. On average, raters who self-identified as Black assigned face images lower ratings than raters of other races (Figure 11B). The magnitude of this rater race effect was greatest when comparing raters who self-identified as Black to raters who self-identified as White, corresponding to an increase of 2.1 $L^*$ units on the CST scale and 2.7 $L^*$ units on the MST scale (Table 3, L* Ratio).

Table 3. Study 2 Linear Mixed Effect Modeling Results for Each Scale

| Scale | | Covariate | Estimate | Standard Error | Profile 95% Confidence Interval | L* Ratio | Conditional $R^2$ |
|---|---|---|---|---|---|---|---|
| CST | Subject | Calibrated L* | -0.1640 | 0.0026 | (-0.1697, -0.1583) | 1.0000 | 0.8873 |
| | | Hue | 0.0159 | 0.0026 | (0.0101, 0.0217) | -0.0968 | |
| | | Chromaticity | -0.0439 | 0.0043 | (-0.0534, -0.0344) | 0.2679 | |
| | | White | -2.1827 | 0.0655 | (-2.3281, -2.0371) | 13.3114 | |
| | | Male | -0.1304 | 0.0360 | (-0.2104, -0.0505) | 0.7955 | |
| | Rater | Asian | 0.2907 | 0.0402 | (0.212, 0.3695) | -1.7731 | |
| | | Hispanic | 0.2956 | 0.0368 | (0.2235, 0.3677) | -1.8028 | |
| | | White | 0.3467 | 0.0376 | (0.273, 0.4203) | -2.1141 | |
| | | Male | -0.0794 | 0.0257 | (-0.1298, -0.029) | 0.4843 | |
| | Device | D | -0.2960 | 0.0312 | (-0.3572, -0.2349) | 1.8055 | |
| | | E | 0.6453 | 0.0312 | (0.5841, 0.7065) | -3.9353 | |
| MST | Subject | Calibrated L* | -0.1173 | 0.0049 | (-0.128, -0.1065) | 1.0000 | 0.8352 |
| | | Hue | 0.0053 | 0.0049 | (-0.0056, 0.0163) | -0.0455 | |
| | | Chromaticity | -0.0188 | 0.0081 | (-0.0368, -0.0009) | 0.1606 | |
| | | White | -2.0687 | 0.1241 | (-2.3443, -1.793) | 17.6417 | |
| | | Male | -0.1924 | 0.0682 | (-0.3437, -0.041) | 1.6405 | |
| | Rater | Asian | 0.2677 | 0.0383 | (0.1925, 0.3428) | -2.2828 | |
| | | Hispanic | 0.2655 | 0.0372 | (0.1927, 0.3384) | -2.2644 | |
| | | White | 0.3201 | 0.0375 | (0.2465, 0.3936) | -2.7295 | |
| | | Male | -0.0549 | 0.0253 | (-0.1045, -0.0054) | 0.4685 | |
| | Device | D | -0.3198 | 0.0305 | (-0.3796, -0.26) | 2.7271 | |
| | | E | 0.2769 | 0.0305 | (0.2171, 0.3368) | -2.3617 | |

# 4 DISCUSSION

Here we sought to measure the sensitivity, consistency, and colorimetric accuracy of skin tone rating scales and understand the factors that influence these ratings. We tested the performance of the Fitzpatrick Skin Type (FST) and two palette-based scales: the Monk Skin Tone (MST) scale and a new Colorimetric Skin Tone (CST) scale developed based on calibrated skin tone measurements. Overall, our findings show that palette-based scales are more sensitive to skin lightness relative to the FST, the factor of primary interest for skin tone scales. Of the two palette-based scales, we found that the CST was more accurate and consistent both for rating own skin tone and for rating the skin tone of faces in images. However, all scale ratings were biased by multiple factors unrelated to skin tone, including large demographic effects especially pronounced when rating the skin tone of faces in images.

The importance of using phenotypes in understanding differential performance of technology was highlighted by Buolamwini and Gebru, who showed that the performance of gender classification algorithms varied based on FST [4]. Since then, the importance of skin tone as a major factor related to performance has been shown for other technologies, from face recognition [7] to pulse oximetry [18]. Recognizing the flaws in the FST scale [17], the MST was developed to provide a more inclusive palette [16, 33]. However, to improve technologies like biometric cameras and pulse-oximeters, engineers need to know the actual spectral properties of skin tones based on standard measures of color. We found that the MST scale colors do not accurately represent skin color (for lighter skin in particular), and this led us to consider how a Colorimetric Skin Tone scale, developed to represent human skin color in standard CIELAB space, would perform. To our surprise, we found that the CST scale was not only more accurate, but could be preferred relative to the MST, especially by individuals with darker skin tones.

When asked to indicate which scale best represented their skin tone, all rater race groups preferred the CST scale on a gray background. On a white background, race groups with lighter skin tones preferred the MST scale. This finding may be due to color contrast: high-contrast varied color backgrounds can make color discrimination difficult while, a low-contrast uniform color background can improve color discrimination [3]. The white background provides less contrast between the lighter skin tone swatches on the MST scale and the background, which improves the discriminability of these scale colors. This is consistent with our observation that preference for the MST increased systematically for people with lighter skin despite the inconsistency between these color swatches and standardized skin tone measurements. People in photographs are seldom pictured against a white background, we believe that a neutral gray background, therefore, provides a better reference point for rating scales.

The CST scale was specifically designed to span the gamut of human skin lightness in CIELAB space. This led ratings selected on the CST scale to be more sensitive to skin lightness than the MST scale and to better match skin color measured using a calibrated instrument. The degree of color error in CST scale choices was consistent across race categories. The MST scale, however, was less accurate when rating the skin tone of people with lighter skin. These findings were observed both for rating own skin tone (study 1) as well as for rating the skin tone of others in images (study 2).

Despite these results, our findings also highlight major confounds intrinsic to human skin tone rating tasks [5, 33]. First, we found that the environment in which palette-based scales are presented can have a strong effect on scale ratings. Ratings of own skin tone were biased by the background on which the scale was presented in a manner consistent with simultaneous brightness contrast [34]. Scale ratings were also biased by the broader experimental environment, differing systematically between experimental locations. Modelling showed that the magnitude of these effects was comparable to a change in skin lightness of about 3 to 6 $L^*$ units. Whereas the on-screen background of scale presentation can be controlled relatively easily, response variation introduced by varied lighting across locations is likely to affect scale ratings performed in distributed environments.

Second, our work shows that skin tone ratings, especially ratings of skin tone from face images, are strongly affected by race. Raters were biased both by their own race as well as the race of the subject being rated. When rating their own skin tone, raters who self-identified as White selected lower scale values (corresponding to lighter skin tones) than raters in other races even when their measured $L^*$ values were similar. Modeling showed that the magnitude of this race bias was comparable to a change in skin lightness of about 6 to 8 $L^*$ units, a larger effect than the effects of background and test environment.

Race bias was even larger when rating the skin tone of others. On average, raters gave lower ratings to images of subjects who self-identified as White than to those that self-identified as Black. This bias was by far the largest observed in this study, comparable to a change in skin lightness of about 13 to 18 $L^*$ units. In contrast to the suggestion that increasing the number and diversity of raters can overcome bias in skin tone ratings [33], we found that raters of all races were similarly biased and strong bias was observed even with approximately 250 diverse raters per image with each scale. These findings indicate that human skin tone ratings are strongly confounded by perceived race as seen in prior research [2]. Taken together with additional bias observed based on raters' race, we found that, on average, a subject who self-identified as White rated by a rater who self-identified as Black is rated 2.4 to 2.5 scale steps lower than a subject who self-identified as Black rated by a rater who self-identified as White, a difference spanning about a quarter of the 10-point scale range (see Table 3; comparisons to reference categories of self-identified Black subjects and raters). This combined race bias roughly corresponds to 15.4 to 20.3 $L^*$ units, spanning 30-50% of the gamut of observed human skin lightness values. Though biases were observed on both the MST and the CST scales, race biases were smaller, in equivalent $L^*$ units, on the CST scale, consistent with its higher sensitivity to skin lightness compared with other factors.

We did find one result suggesting that human perception may improve accuracy of skin tone estimation from images. The lightness of subjects' skin can vary dramatically across imaging devices, even when taken under similar conditions [17]. When presented with such varying images from different devices, skin tone ratings were more consistent than expected based on the measured lightness of faces in the image. On average, variation in skin tone ratings across imaging devices was equivalent to 5 to 6 $L^*$ units (see Table 3; difference between D and E) compared with 23 $L^*$ unit variation across imaging devices (see Figure 11A, difference between D and E). This suggests that raters can analyze the image to correct their skin tone estimates for over or under-exposure. We believe that replicating this ability in AI-based image analysis systems, while avoiding race-bias, holds promise for improving automated methods of skin tone estimation [12].

Although human annotation of skin tone may never be objective, the process can be improved by using the CST scale. Using color values directly traceable to calibrated skin tone measurements from a diverse population, we show that this scale is inclusive and offers an advance over prior art. When used appropriately, it provides a more sensitive, consistent, and accurate method of assessing variation in human skin lightness for performing technology evaluations and informing engineering improvements needed to build more inclusive technologies.

## ACKNOWLEDGMENTS

This research was sponsored by the United States Department of Homeland Security's Science and Technology Directorate on contract numbers 70RSAT18CB00000034 and 70RSAT23CB00000003. Author contributions: CMC selected and guided statistical analysis methods, wrote, and edited the paper; JJH conceived the work and supported the development of the CST scale, directed statistical analyses, and edited the paper; LRR conceived the work, designed survey instruments, and administered surveys; IMS designed survey instruments, administered the surveys, performed statistical analyses, wrote and edited the paper; YBS conceived and directed the work, developed the CST scale, designed survey instruments, directed statistical analyses, wrote and edited the paper; JLT conceived the work and edited the paper; ARV conceived the work and edited the paper. Thanks to the staff of the SAIC Identity and Data Sciences Laboratory without whom this work would not be possible. Special thanks to Dr. Trey Wood and Dr. Richard Plesh of the SAIC Identity and Data Sciences Laboratory for careful reading and helpful comments on the manuscript and to Dr. Daniela Barragan for initial exploratory data analyses.